\documentclass{iau} 
\usepackage{amsmath,amssymb,revsymb,graphicx,dcolumn}
\usepackage{leftidx}
\usepackage{slashed}

\newcommand{\beq}{\begin{equation}}
\newcommand{\eeq}{\end{equation}}
\newcommand{\beqa}{\begin{eqnarray}}
\newcommand{\eeqa}{\end{eqnarray}}
\newcommand{\bsubeqs}{\begin{subequations}}
\newcommand{\esubeqs}{\end{subequations}}

\newcommand{\imineq}[2]{\vcenter{\hbox{\includegraphics[height=#2ex]{#1}}}}

\title[QED far from evaporating black holes] 
{Quantum electromagnetic phenomena \\ far from small evaporating black holes}

\author[Slava Emelyanov]   
{Slava Emelyanov}

\affiliation{Institute of Theoretical Physics, Karlsruhe Institute of Technology (KIT), \\ 76128 Karlsruhe,
Germany \\ email: {\tt viacheslav.emelyanov@kit.edu}}

\pubyear{2016}
\volume{324}  
\jname{New Frontiers in Black Hole Astrophysics}
\editors{A.C. Editor, B.D. Editor \& C.E. Editor, eds.}
\begin{document}

\maketitle

\begin{abstract}
One might expect far away from physical black holes that quantum field quantisation performed in Minkowski
space is a good approximation. Indeed, all experimental tests in particle colliders reveal no deviations so far.
Nevertheless, the black holes should leave certain imprints of their presence in quantum processes. In this 
paper, we shall discuss several local imprints of small, primordial evaporating black holes in quantum electrodynamics 
in the weak gravity regime. Physically this can be interpreted as being macroscopic manifestations of vacuum
fluctuations.
\end{abstract}

\firstsection 
\section{Introduction}
In quantum field theory observables are self-adjoint, local operators. These local operators
could be of various nature. In what follows we consider observables corresponding to the
electromagnetic field. Specifically, these could be an electromagnetic wave propagating 
from an emitter to a receiver and an electrostatic field sourced by a point-like charge, both
being studied in a box of about one-cubic-meter size.

It is a consequence of the requirements needed to formulate a physically acceptable quantum
field theory (e.g., \cite[Haag 1996]{Haag}) that quantum operators localised in a certain space-time
region probe the vacuum as if it is a non-empty state. The physical interpretation of this effect 
can be readily given in terms of the quantum or vacuum fluctuations. These fluctuations or
quantum noise can show up at microscopic as well as macroscopic scales. The formation of 
inhomogeneities like galaxies and clusters of galaxies in our universe is their most remarkable
macroscopic manifestation (\cite[Mukhanov \& Chibisov 1981]{Mukhanov&Chibisov}). 

According to the Tomita-Takesaki theorem (\cite[Haag 1996]{Haag}), the quantum vacuum
satisfies a Kubo-Martin-Schwinger aka thermal condition with respect to any localised set
of the field operators and a certain one-parametric group of its automorphism. This group
could be of a geometrical as well as non-geometric origin, i.e. generated by a Killing or 
non-Killing vector, respectively. The lesson of this is that the quantum vacuum can respond
as a many-particle state which is mainly a property of quantum field operators, rather than
of the vacuum state.

\firstsection

\section{Quantum vacuum as plasma-like medium}

{\underline{\it Vacuum as dielectric}}. The quantum vacuum possesses a \emph{dielectric}-like
property in the absence of any external field with one of its observable/measurable manifestation
being a part of the Lamb shift. The ``dipoles" of the vacuum are virtual $e^+e^-$ pairs. This effect
follows from taking into account the one-loop correction to the photon self-energy which leads to
the running electric charge (e.g., \cite[Peskin \& Schroeder 1995]{Peskin&Schroeder}). 

Specifically, the full photon propagator in quantum field theory is diagrammatically given by
\beqa
\mathbf{\imineq{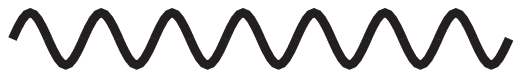}{2.0}}+\mathbf{\imineq{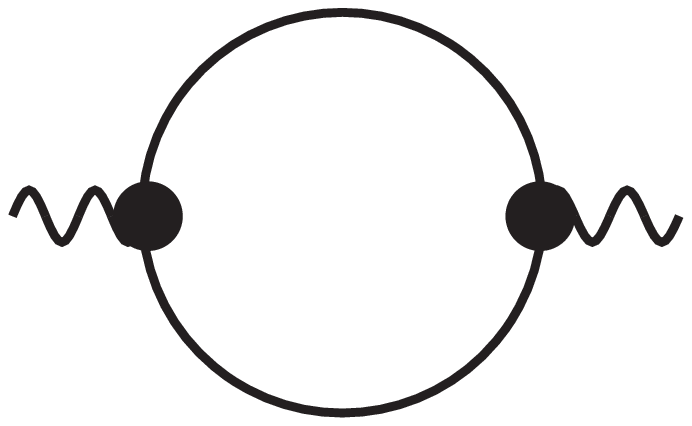}{8.4}} &+&
\text{higher-loop diagrams}\,.
\eeqa
The loop diagrams lead to the modification of the classical field equation. This becomes in momentum 
space
\beqa
\Big(k^2\eta^{\mu\nu} - k^\mu k^\nu - \Pi^{\mu\nu}(k)\Big)a_\nu(k) &=& 0\,,
\eeqa
where $\Pi^{\mu\nu}(k)$ is a polarisation tensor vanishing in the limit of the vanishing fine structure 
constant $\alpha$. The gauge symmetry entails 
$\Pi^{\mu\nu}(k) = (k^2\eta^{\mu\nu} - k^\mu k^\nu)\Pi(k^2)$. Thus, one obtains 
$k^2(1 - \Pi(k^2))a^\mu = 0$ for the photon ($k^\mu a_\mu = 0$). The extra factor $1 - \Pi(k^2)$ leads
to the charge renormalisation and, hence, to the modification of the classical Coulomb interaction
at scales of the order of the Compton length of the electron. The resulting effect is the screening of the
charge by the virtual $e^+e^-$ pairs like in a dielectric.

{\underline{\it Vacuum as magnetised plasma}}. Probing the quantum vacuum by studying how an
electromagnetic wave propagates through a spacetime region with a super-strong magnetic field
($B \gg (\pi/\alpha)B_c$, where $B_c = m_e^2/e$), the vacuum shows up a property inherent to a 
magnetised plasma, i.e. the plasma held at a static external magnetic
field (\cite[Melrose \& Stoneham 1976]{Melrose&Stoneham} and \cite[Dittrich \& Gies 2000]{Dittrich&Gies}).

Due to a collective response of the plasma particles being tied to the external magnetic field lines, the 
propagation of an electromagnetic wave is significantly dependent on its polarisation
(e.g., \cite{Bittencourt}).
Specifically, an electromagnetic wave polarised in the direction coplanar to the plane defined by its 
spreading direction $\mathbf{k}$ and the external magnetic field $\mathbf{B}$ propagates along 
the magnetic field lines. This is known as the Alfv\'{e}n wave.

The external magnetic field $B$ in the vacuum entails the modification of the vacuum polarisation
tensor $\Pi^{\mu\nu}(k,B)$ which becomes non-diagonal. This is due to $B \neq 0$ which enters
the field equations and, hence, modifies the electron and photon propagator. At one-loop approximation
it turns out that a light wave polarised in the direction perpendicular to the plane defined by $\mathbf{k}$ 
and $\mathbf{B}$ is oblivious to the external magnetic field, i.e. $\omega^2 = |\mathbf{k}|^2$.
This resembles a behaviour of the fast magneto-acoustic wave in case of the Alfv\'{e}n velocity
$v_A = |\mathbf{B}|/\sqrt{4\pi\rho} \gg 1$,
where $\rho$ is a plasma-like density of the vacuum. If the polarisation vector $a_\mu$ is coplanar
to that plane, then, as it was found by~\cite[Melrose \& Stoneham (1976)]{Melrose&Stoneham},
$\omega^2 = |\mathbf{k}|^2\cos^2\theta$, 
where $\theta$ is an angle between $\mathbf{k}$ and $\mathbf{B}$. This corresponds to the Alfv\'{e}n 
wave in case the condition $v_A \gg 1$ is fulfilled.

{\underline{\it Vacuum as neutral plasma}}. The quantum vacuum responds non-trivially in the
background of evaporating black holes, namely it acquires plasma-like 
properties (\cite[Emelyanov 2016a]{Emelyanov-2016-2}).

To demonstrate the effect, one needs to compute the vacuum polarisation tensor. It must depend \textit{a priori}
on a mass $M$ of the hole and a distance $R$ to its centre. The former dependence arises since
when $M = 0$ the problem reduces to that in flat space-time. The latter one is, however, due to 
our expectation that \emph{local} physics in the asymptotically flat region ($R \rightarrow \infty$) should
reduce to the familiar Minkowskian one.

A part of the tensor $\Pi^{\mu\nu}(k,M,R)$ was computed at one-loop level 
by~\cite[Drummond \& Hathrell (1980)]{Drummond&Hathrell}. However, it gets an extra term due to
the quantum effect entailing the black-hole evaporation. Specifically for the semi-classical theory be
reliable, a field operator in the background of any black hole should be expanded over modes in which
the vacuum expectation value of stress tensor does not diverge on the (future) horizon. This yields
the electron propagator in momentum space of the following form
\beqa\label{eq:e-propagator-bh}
S(k,k') &\approx& \big(\slashed{k} + m_e\big)\left(\frac{i}{k^2 - m_e^2 +i\varepsilon} 
- 2\pi\Big(\frac{27r_H^2}{16R^2}\Big)\frac{\delta(k^2 - m_e^2)}{e^{k_0/T_H} + 1}\right)\delta(k-k')\,,
\eeqa
where $T_H = M_\text{Pl}^2/8\pi M$ is Hawking's parameter, $M_\text{Pl} = (\hbar/G)^\frac{1}{2}$
the Planck mass, $r_H \equiv 2MG$ a size of the black-hole horizon. We have omitted 
in~\eqref{eq:e-propagator-bh} terms of sub-leading order vanishing as $1/R^3$ at the spatial infinity.
The propagator~\eqref{eq:e-propagator-bh} is valid only for $|r-r'| \ll R$  and
$r_H \ll R$ (\cite[Emelyanov 2016a \& 2016b]{Emelyanov-2016-2,Emelyanov-2016-3}). We should 
mention that it is legitimate in this regime to employ flat coordinates as space-time is \emph{locally}
Minkowski.

The geometrical part of $\Pi^{\mu\nu}(k,M,R)$ drops out at least as fast as $1/R^3$ at spatial infinity,
whereas its vacuum part vanishes as $1/R^2$ as can be seen in~\eqref{eq:e-propagator-bh}. Therefore,
the main contribution to $\Pi^{\mu\nu}(k,M,R)$ is due to the second term in~\eqref{eq:e-propagator-bh}.

The modification of the electron propagator has a thermal-like structure. In case of a large black hole
($m_e \gg T_H$) the one-loop self-energy of the photon is exponentially suppressed by the Boltzmann
factor $\exp(-m_e/T_H)$. Therefore, we consider a small black hole with $T_H \gg m_e$ in the following.
This corresponds to the black-hole mass $M \ll 10^{16}\,\text{g}$ and, thus, justifies the hard thermal
loop approximation (e.g., \cite{LeBellac}) in which one can omit the electron mass $m_e$ as being
negligible with respect to the temperature parameter $T_H$.

At one-loop level the photon propagator acquires two poles. One of these corresponds to a transverse
mode, the photon, with $\omega^2 = |\mathbf{k}|^2 + \pi_T(\omega,|\mathbf{k}|)$, another one to a
longitudinal mode with $\omega^2 = |\mathbf{k}|^2 + \pi_L(\omega,|\mathbf{k}|)$ which is known as 
a plasmon. The functions $\pi_T(\omega,|\mathbf{k}|)$ and $\pi_L(\omega,|\mathbf{k}|)$ are the 
same as found by~\cite[Weldon (1982)]{Weldon}, but with the temperature
\beqa
T_L &\approx& \frac{3\sqrt{3}}{16\pi}M_\text{Pl}\frac{L_\text{Pl}}{R}\,, 
\eeqa
where $L_\text{Pl} = (\hbar G)^{\frac{1}{2}}$ is the Planck
length (\cite[Emelyanov 2016a]{Emelyanov-2016-2}). In the 
leading-order approximation $T_L$ does not directly depend on the mass $M$ of the black
hole. The quantum vacuum unlike a physical hot plasma described by $\alpha$ and $T$ (but not,
e.g., by a particle density $n$) is effectively also characterised by $T_L$ which unlike $T_H$ plays
a major role in local physics.

The wavelength of the mode has to be much smaller than the distance to the black hole. Moreover, it
must be much smaller than a detector size $l_D$. This implies 
$m_e^2/eT_H  \gg |\mathbf{k}|$ and $|\mathbf{k}| \gg 1/l_D \gg 10^2\omega_\text{p}$,
where $\omega_\text{p} \equiv \frac{1}{3}\,eT_L$ 
is a plasma-like frequency (\cite[Emelyanov 2016a]{Emelyanov-2016-2}) and the upper threshold is 
to suppress higher loop corrections (\cite[Emelyanov 2016c \& 2016a]{Emelyanov-2016-1,Emelyanov-2016-2}).
In this case, plasmons are not propagating modes (\cite{LeBellac}), whereas photons acquire an effective mass
\beqa
m_\gamma &\approx& \frac{1}{\sqrt{6}}\,eT_L \;\approx\; 2.5{\times}10^{-9}
\left(\frac{1\,\text{m}}{R}\right)\,\text{eV}\,.
\eeqa
In the physical plasma, plasmons are quasi-particles, i.e. collective excitations mediated by plasma
particles. In our case, the vacuum is full of \emph{virtual} particles, rather than real ones. 
Thus, the result of having no plasmons seems to be physically sensible.

A static electric field $\mathbf{E}$ sourced by a point-like charge $q$ is screened in the physical 
plasma - a phenomenon known as the Debye screening. This effect is due to a collective response 
of the plasma particles on putting the charge $q$ in the system. The linear response theory in our
case yields a similar result, namely $\mathbf{E} = -\nabla\varphi$ with the electromagnetic potential
$\varphi = (q/r)\,e^{-r/r_L}$, where
\beqa
r_L &=& \big(\sqrt{2}m_\gamma\big)^{-1} \;\approx\; 3.5{\times}10^2R\,.
\eeqa
Thus, the Debye-like length $r_L$ is much bigger than $R$. It means the screening is a tiny effect,
because $l_D \ll R$ should hold for our approximation be reliable (at distances $|r-r'| \gtrsim R$
the propagator \eqref{eq:e-propagator-bh} is invalid). With the technology
of~\cite[Williams \etal\ (1971)]{Williams&Faller&Hill} this can in principle be tested if such small
black holes do exist in nature and one of them is within $R \lesssim 250\,\text{km}$ from the 
detector. For a practical purpose, however, the probability of having such a black hole per cubic
centimeter should be estimated to decide whether it is realistic to probe this effect during a
reasonable amount of time.

If the black-hole evaporation lasts till its complete disappearance, then we should set a lower bound on
the black-hole mass $M$ to guarantee the quasi-equilibrium approximation we have exploited above. 
It would be of order of $10^{10}\,\text{g}$ corresponding to the black-hole lifetime of one day.
The \emph{initial} mass of this primordial hole (i.e.~$10^{10}\,\text{years}$ ago) should~be in the
range $3{\times}10^{14}\,\text{g} \lesssim M_0 \ll 10^{16}\,\text{g}$, which corresponds to 
$10^{10}\,\text{g} \lesssim M \ll 10^{16}\,\text{g}$ today.

\firstsection

\section{Discussions}

We have considered three examples when the quantum vacuum can be envisaged as possessing
medium-like properties. The \emph{virtual} particles were understood as constituents of the vacuum
like the \emph{real} particles of ordinary media - the point of view which seems to be widely-accepted 
by the researchers.
 
As with any analogy, this viewpoint cannot be taken literally. Indeed, the real particles can transfer
energy, whereas the virtual particles seem not be capable of doing that, because these are
just fluctuations, although quantum in nature.

To clarify this aspect of the quantum fluctuations, one may put an extra conducting plate between the
Casimir plates. It may be taken to be freely moving and at the distance $l_C/2$ from each
plates at $t < 0$. Its motion is however fixed by $v(t) = v\,\theta(t)\theta(t -~\tau) \ll 1$. It 
follows from energy conservation that a part of the vacuum energy in both cavities will by absorbed
by the middle plate during its motion. In other words, the vacuum energy is partially dissipated in the 
middle plate and partially redistributed between the cavities. The dissipative part could result in heating
of the middle plate with an increase in temperature of $\Delta T \propto V(v\tau)^2/l_C^6$ in case
$v\tau \ll l_C$, where $V$ is its volume.~This \emph{gedankenexperiment} might imply the virtual
particles are capable of transferring energy.

\firstsection

\section*{Acknowledgements}

I am grateful to the organisers of IAU Symposium 324 for the financial support. It is a pleasure
to thank the anonymous referee for his physically relevant comments/questions.

\firstsection

\end{document}